\documentclass[%
 reprint,
superscriptaddress,
showpacs,
 amsmath,amssymb,
 aps,
prb,
floatfix, longbibliography ]{revtex4-1}

\usepackage{graphicx}
\usepackage{dcolumn}
\usepackage{bm}
\usepackage{natbib}
\usepackage{color}
\usepackage{soul}
\usepackage{amssymb,amsmath}

\begin{document}

\title{Origin of Unusually High Rigidity in Selected Helical Coil Structures}

\author{David Tom\'{a}nek}
\email
            {tomanek@pa.msu.edu}%
\affiliation{Physics and Astronomy Department,
             Michigan State University,
             East Lansing, Michigan 48824, USA}

\author{Arthur~G.~Every}
\affiliation{School of Physics,
             University of the Witwatersrand,
             Private Bag 3,
             2050 Johannesburg,
             South Africa
}

\date{\today} 

\begin{abstract}
Using continuum elasticity theory, we describe the elastic
behavior of helical coils with an asymmetric double-helix
structure and identify conditions, under which they become very
rigid. Theoretical insight gained for macro-structures including a
stretched telephone cord and an unsupported helical staircase is
universal and of interest for the elastic behavior of helical
structures on the micro- and nanometer scale.
\end{abstract}



\maketitle

\section{Introduction}

\begin{figure}[t]
\includegraphics[width=1.0\columnwidth]{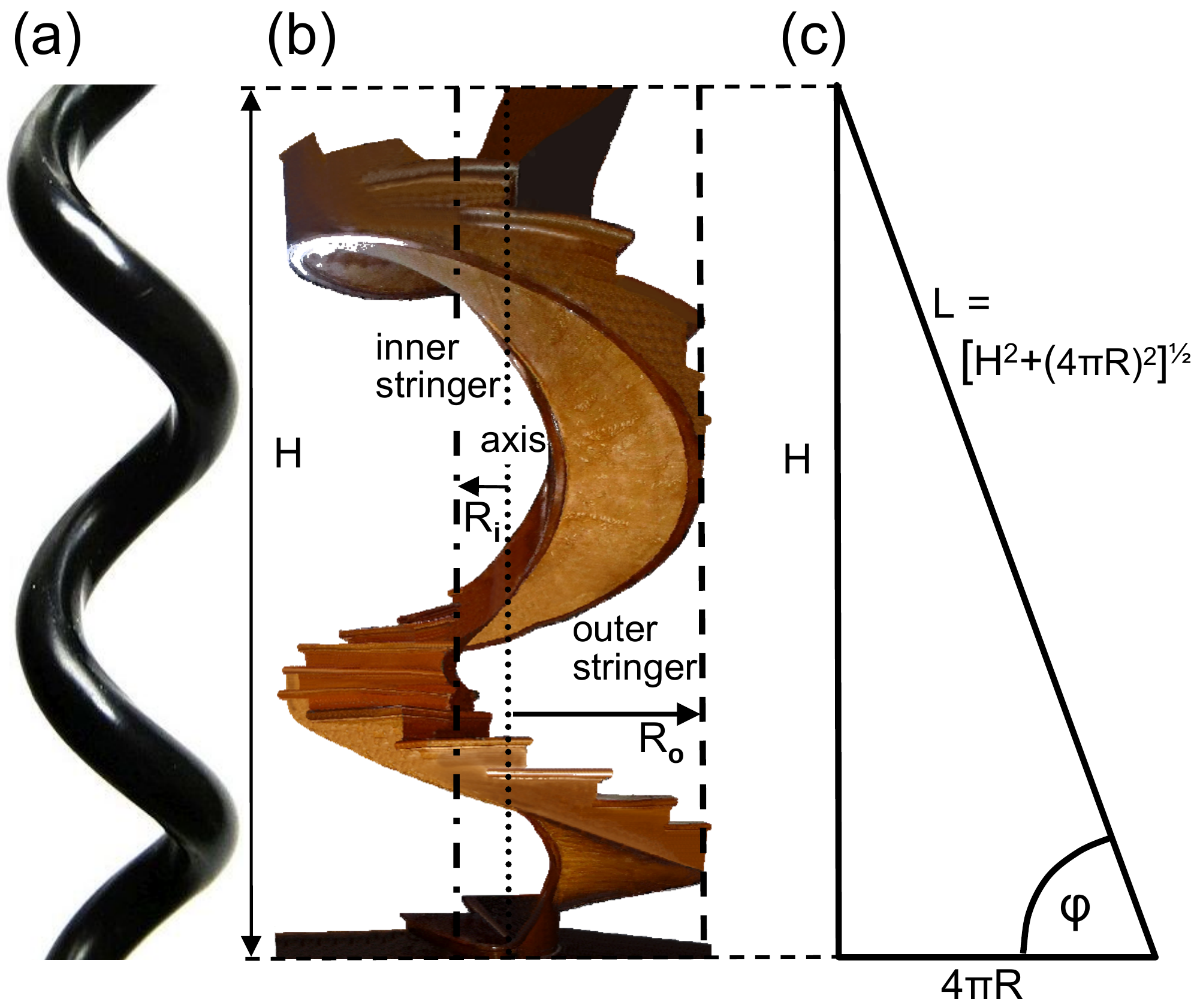}
\caption{(a) Photograph of a coiled telephone cord with the same
topology as an unsupported spiral staircase. (b) Retouched
photograph of the unsupported spiral staircase at the Loretto
Chapel as constructed. (c) Trace of the helical inner or outer
stringers of the staircase on the surface of a cylinder, which can
be unwrapped into a rectangle. \label{fig1} }
\end{figure}

Helical coil structures, ranging from a stretched telephone cord
in Fig.~\ref{fig1}(a) and an unsupported spiral staircase in
Fig.~\ref{fig1}(b) on the macro-scale to DNA and proteins on the
micro-scale abound in Nature. Since their elastic behavior is
governed by the same laws of Physics independent of scale, insight
obtained on the macro-scale will benefit the understanding of
helical micro- and nanostructures. An intriguing example of
unusual high rigidity on the macro-scale, which has remained
unexplained to date, is the unsupported all-wooden spiral
staircase in the Loretto Chapel~\cite{Loretto-Chapel-2002} in
Santa Fe, New Mexico, constructed around 1878 and shown in
Fig.~\ref{fig1}(b). In the following we explore the elastic
behavior of this structure using continuum
elasticity theory in order to identify the reason for its rigidity~\cite{%
{Love-book},{Fakhreddine05},{Zubov07}}.
Since continuum elasticity theory applies from nanometer-sized
fullerenes and nanotubes~\cite{{DT071},{DT255},{DT260}} to the
macro-scale, we expect our approach to be useful to explore the
rigidity of helical structures on the micro- and nanometer scale.

The use of continuum elasticity theory rather than the
case-specific finite-element method~\cite{Fakhreddine05} in this
case is motivated by our objective to identify the universal
origin of the high rigidity of the Loretto spiral staircase and
related helical coils with an asymmetric double-helix structure.
Theoretical insight gained for macro-structures including a
stretched telephone cord and an unsupported helical staircase is
universal and of interest for the elastic behavior of helical
structures on the micro- and nanometer scale.

\section{Elastic Behavior of a Helical Coil}

From a Physics viewpoint, the spiral staircase of
Fig.~\ref{fig1}(b) is a compression coil or a helical spring with
a rather high pitch. It can be characterized as an asymmetric
double-helix structure consisting of an inner stringer coil of
radius $R_i$ and an outer stringer coil of radius $R_o$, and spans
two turns in total. The two stringer coils are connected by rigid
steps of width $R_o-R_i$. The staircase can be equivalently
described as a helicoid, or a ``filled-in'' helix, with finite
nonzero inner and outer radii.

There is an extensive literature on the properties of helical
springs, which dates back to Love's treatise~\cite{Love-book}, as
well as the more recent
Refs.~[\onlinecite{Fakhreddine05},\onlinecite{Zubov07}] and
literature cited therein. Yet the compound helical structure of
the Loretto staircase and related helical coils appears to have
escaped attention in publications so far. The rigidity of the
connecting steps provides the spiral staircase with a remarkable
degree of stiffness, as we show below. We propose that this
property is shared by similar helical structures independent of
their scale.

Any coil of radius $R$ and total height $H$, such as the inner and
the outer stringer of the staircase, lies within a cylindrical
wall of the same height, which can be unwrapped onto a triangle,
as shown in Fig.~\ref{fig1}(c). Let us first consider the case of
incompressible inner and outer stringers of height $H$ that are
separated by the constant distance $R_o-R_i$, which defines the
step width.

According to Fig.~\ref{fig1}(c), the equilibrium length $L_i$ of
the inner stringer coil with two turns is related to its
equilibrium radius $R_i$ and its equilibrium height $H$ by
\begin{equation}
L_i^2 = H^2 + (4{\pi})^2R_i^2 \,. %
\label{eq1}
\end{equation}
The equivalent relation applies, of course, to the outer stringer
of radius $R_o$.


We first consider an incompressible stringer of constant length
$L_i$, which is stretched axially by ${\delta}H$, causing the
radius $R_i$ to change by ${\delta}R_i$. Then,
\begin{equation}
L_i^2 = (H+{\delta}H)^2 + 16{\pi}^2(R_i+{\delta}R_i)^2 \,. %
\label{eq2}
\end{equation}
Subtracting Eq.~(\ref{eq1}) from Eq.~(\ref{eq2}) and ignoring
${\delta}H^2$ and ${\delta}R_i^2$ terms in the limit of small
deformations, we obtain
\begin{equation}
2H{\delta}H + 16{\pi}^2(2R_i{\delta}R_i)  = 0 %
\label{eq3}
\end{equation}
and consequently
\begin{equation}
{\delta}R_i = -\frac{H}{16{\pi}^2R_i} {\delta}H \,. %
\label{eq4}
\end{equation}
Considering the outer helical stringer to behave independently for
the moment, we expect the counterpart of Eq.~(\ref{eq4})
\begin{equation}
{\delta}R_o = -\frac{H}{16{\pi}^2R_o} {\delta}H %
\label{eq5}
\end{equation}
to describe the outer stringer. The step width should then change
by
\begin{equation}
{\delta}(R_o - R_i) = %
{\delta}R_o - {\delta}R_i = %
-\frac{H{\delta}H}{16{\pi}^2} %
\left( \frac{1}{R_o} - \frac{1}{R_i} \right) \,. %
\label{eq6}
\end{equation}
The only way to keep the step width constant, corresponding to
${\delta}(R_o-R_i)=0$, is to have either zero step width
$R_i=R_o$, reducing the double-helix to a single-helix, or to
suppress any change in height with ${\delta}H=0$. Even though each
individual stringer coil can change its height $H$ while keeping
its length $L$ constant, the assumed rigid connection between the
inner and outer stringers makes the staircase completely rigid.


Next, we relax the constraint that each stringer should maintain
its length when the staircase changes its height $H$.
Nevertheless, we will still maintain the assumption of a fixed
step width
\begin{equation}
R_o-R_i = const. %
\label{eq7}
\end{equation}
that translates to ${\delta}R_i={\delta}R_o={\delta}R$. The
compressible inner stringer helix will still be characterized by
its equilibrium length $L_i$ and equilibrium radius $R_i$. Its
deformation caused by changes of its axial height by ${\delta}H$
is then described by
\begin{equation}
(L_i+{\delta}L_i)^2 %
= (H+{\delta}H)^2 + (4{\pi})^2(R_i+{\delta}R_i)^2 \,. %
\label{eq8}
\end{equation}
Ignoring ${\delta}L_i^2$, ${\delta}H^2$ and ${\delta}R_i^2$ terms
in the limit of small deformations, we obtain
\begin{equation}
L_i^2+2L_i{\delta}L_i %
= H^2+2H{\delta}H + 16{\pi}^2(R_i^2+2R_i{\delta}R_i) \,. %
\label{eq9}
\end{equation}
Subtracting Eq.~(\ref{eq1}) from Eq.~(\ref{eq9}), we obtain
\begin{equation}
2L_i{\delta}L_i = 2H{\delta}H + 16{\pi}^2 2R_i{\delta}R_i \,. %
\label{eq10}
\end{equation}
With the assumption ${\delta}R_i={\delta}R_o={\delta}R$, we can
rewrite Eq.~(\ref{eq10}) and its counterpart for the outer
stringer as
\begin{eqnarray}
L_i{\delta}L_i = H{\delta}H + 16{\pi}^2 R_i{\delta}R \,, \nonumber \\%
L_o{\delta}L_o = H{\delta}H + 16{\pi}^2 R_o{\delta}R \,.
\label{eq11}
\end{eqnarray}
Combining all terms containing ${\delta}R$ on one side, we can
eliminate ${\delta}R$ by dividing the two equations. This leads to
\begin{equation}
\frac{L_i{\delta}L_i-H{\delta}H}{L_o{\delta}L_o-H{\delta}H} %
= \frac{R_i}{R_o} %
\label{eq12}
\end{equation}
and, by rearranging terms, to
\begin{equation}
\frac{L_i}{R_i}{\delta}L_i - \frac{H}{R_i}{\delta}H = %
\frac{L_o}{R_o}{\delta}L_o - \frac{H}{R_o}{\delta}H = \kappa \,,
\label{eq13}
\end{equation}
where $\kappa$ is a variable to be determined by minimizing the
strain energy $U$ of the deformed stringers. $U$ is given by
\begin{eqnarray}
U &=& %
  \frac{1}{2}C \left(%
  \frac{{\delta}L_i}{L_i} \right)^2 {L_i} %
+ \frac{1}{2}C \left(%
  \frac{{\delta}L_o}{L_o} \right)^2 {L_o} \nonumber \\%
&=& \frac{1}{2}C \left(%
  \frac{{\delta}L_i^2}{L_i} + \frac{{\delta}L_o^2}{L_o} %
  \right) \,,
\label{eq14}
\end{eqnarray}
where $C$ is the force constant describing the elastic response of
the stringers to stretching. For a macroscopic stringer with the
Young's modulus $E$ and the cross-sectional area $A$, $C=EA$. From
Eq.~(\ref{eq13}) we get
%
%
%
\begin{equation}
{\delta}L_i = \frac{H{\delta}H+{\kappa}R_i}{L_i} %
\label{eq15}
\end{equation}
for the inner stringer. Similarly, we get
\begin{equation}
{\delta}L_o = \frac{H{\delta}H+{\kappa}R_o}{L_o} %
\label{eq16}
\end{equation}
for the outer stringer and can now rewrite the strain energy as
\begin{equation}
U = %
\frac{1}{2}C \left(%
  \frac{(H{\delta}H+{\kappa}R_i)^2}{L_i^3} %
+ \frac{(H{\delta}H+{\kappa}R_o)^2}{L_o^3} %
  \right) \,.
\label{eq17}
\end{equation}
The optimum value of $\kappa$ is obtained from requiring
${\partial}U/{\partial}{\kappa}=0$. This leads to
\begin{equation}
  \frac{(H{\delta}H+{\kappa}R_i)R_i}{L_i^3} %
+ \frac{(H{\delta}H+{\kappa}R_o)R_o}{L_o^3} = 0 \,, %
\label{eq18}
\end{equation}
which can be rewritten as
\begin{equation}
  (H{\delta}H+{\kappa}R_i)R_iL_o^3 %
+ (H{\delta}H+{\kappa}R_o)R_oL_i^3 = 0 \,. %
\label{eq19}
\end{equation}
We can regroup the terms to get
\begin{equation}
H{\delta}H \left( R_i L_o^3 + R_o L_i^3 \right) + %
{\kappa} \left( R_i^2 L_o^3 + R_o^2 L_i^3 \right) = 0 \,,
\label{eq20}
\end{equation}
which yields expressions for the optimum values of $\kappa$,
${\delta}L_i$ and ${\delta}L_o$. We get
\begin{equation}
{\kappa} = -H{\delta}H %
\frac{R_i L_o^3 + R_o L_i^3}{R_i^2 L_o^3 + R_o^2 L_i^3} \,,
\label{eq21}
\end{equation}
\begin{eqnarray}
{\delta}L_i &=& \frac{H{\delta}H}{L_i} %
\left[ 1 - R_i\left( %
\frac{R_i L_o^3 + R_o L_i^3}{R_i^2 L_o^3 + R_o^2 L_i^3} %
\right) \right] \nonumber \\
&=& H{\delta}H \left[ %
\frac{L_i^2 R_o}{R_i^2 L_o^3 + R_o^2 L_i^3}
\right] (R_o-R_i) 
\,, %
\label{eq22}
\end{eqnarray}
and
\begin{eqnarray}
{\delta}L_o &=& \frac{H{\delta}H}{L_o} %
\left[ 1 - R_o\left( %
\frac{R_i L_o^3 + R_o L_i^3}{R_i^2 L_o^3 + R_o^2 L_i^3} %
\right) \right] \nonumber \\
&=& H{\delta}H \left[ %
\frac{L_o^2 R_i}{R_i^2 L_o^3 + R_o^2 L_i^3}
\right] (R_i-R_o) 
\,. %
\label{eq23}
\end{eqnarray}
%
%
To interpret this result, let us first consider the inner and
outer stringers to be independent first and only then consider the
effect of a constant step width separating them. In response to
${\delta}H>0$, the inner stringer prefers to reduce its radius
significantly, but this reduction is limited by the
constant-step-width constraint. Thus, the length of the inner
stringer is increased and it is in tension. For this to occur, the
stairs must have been pulling the inner stringer outwards, and so
the steps are subject to tensile stress. In response to increasing
its height, also the outer stringer prefers to reduce its radius.
But the constant-step-width constraint reduces its radius even
more, so that the outer stringer ends up in compression. To
accomplish this, the steps must be pulling it inward and again
should be subjected to tensile stress. In response to
${\delta}H<0$, the strains in the inner and the outer stringers
will change sign and the steps will be under compressive stress.

The total strain energy amounts to
\begin{equation}
U = \frac{1}{2} C %
\frac{H^2 {\delta}H^2 (R_o-R_i)^2}{R_i^2 L_o^3 + R_o^2 L_i^3} %
= \frac{1}{2} k {\delta}H^2 \,, %
\label{eq24}
\end{equation}
where $k$ is the spring constant of the entire double-helix
structure, given by
\begin{equation}
k = C\frac{H^2 (R_o-R_i)^2}{R_i^2 L_o^3 + R_o^2 L_i^3} \,. %
\label{eq25}
\end{equation}
We note that in in a single-stringer case, characterized by
$R_o-R_i=0$, the axial spring constant $k$ would vanish in our
model.


For the initially mentioned spiral staircase in the Loretto
chapel, $R_i=0.26$~m, $R_o=1.00$~m, and $H=6.10$~m. From
Eq.~(\ref{eq1}), we get $L_i=6.92$~m and $L_o=13.97$~m.

For the sake of a fair comparison to a straight staircase with a
slope given by $\tan(\varphi)$, as seen in Fig.~\ref{fig1}(c), we
do not use the pitch, but rather the local slope
$\tan(\varphi_i)=H/(4{\pi}R_i)$ of the inner stringer to
characterize how steep the staircase is. The Loretto staircase is
rather steep near the inner stringer with $\tan(\varphi_i)=1.9$,
corresponding to ${\varphi_i}{\approx}61^\circ$.

The stringers of the Loretto staircase have a rectangular
cross-section of $6.4$~cm${\times}19.0$~cm, and so we have for the
cross-section area $A=121$~cm$^2$. Considering the elastic modulus
$E{\approx}10^{10}$~N/m$^2$ for wood along the grains, we obtain
$C=E{\cdot}A=1.2{\times}10^8$~J/m. Thus, the spring constant of
the double-helix structure describing the staircase could be as
high as $k=4.8{\times}10^6$~N/m.

Now consider the staircase suspended at the top and free to deform
in the axial direction. The largest deformation will occur when a
load is applied on the lowest step. A person of $100$~kg in that
location would apply net force $F=981$~N to the staircase, causing
an axial elongation of ${\delta}H=F/k=0.2$~mm, which is very
small.

In reality, the staircase is anchored both at the top and the
bottom, and its total height is constrained. The weight of a
person climbing up the stairs is supported by the fraction $x$ of
the staircase below, which is under compression, and the fraction
$(1-x)$ of the staircase above, which is under tension. The local
axial deflection ${\delta}h$ along the staircase is then given by
\begin{equation}
{\delta}h(x) = \frac{F}{k} x(1-x) \,. %
\label{eq26}
\end{equation}
The largest deflection occurs in the mid-point of the staircase,
with $x(1-x)=1/4$. The local vertical deflection caused by a
person of $100$~kg standing at this point should be only
${\delta}h{\approx}0.05$~mm. As expected intuitively, there is no
deflection for a person standing either at the top or at the
bottom.


\section{Bending Deformation of a Helical Coil}

Structurally, the telephone cord in Fig.~\ref{fig1}(a), the
unsupported helical staircase in Fig.~\ref{fig1}(b), and a rubber
hose share one important property: all elastic material is on the
surface of a hollow cylinder, forming a tube. In a further degree
of simplification, we may ignore the interior structure of this
elastic tube and describe its stretching, twisting or bending
deformations using continuum elasticity theory~\cite{DT260}. So
far, we have considered stretching as the dominant response to
tensile stress. %
When a compressive load $F$ is applied to the helical coil, there
will always be a reduction in the height $H$ proportional to $F/H$
due to compression. But there will only be bending, which is
synonymous with buckling, if $FH^2$ exceeds a critical
value{\cite{{Euler-bending},{Timoshenko}}}. Our
task will be to identify this critical value.%

This %
finding %
agrees with published continuum elasticity results for
long-wavelength acoustic phonon modes in tubular
structures~\cite{DT260}, which suggest a fundamentally different
dispersion relation $\omega_{ZA}{\propto}k^2$ for bending modes,
in stark contrast to $\omega_{LA,TA}{\propto}k$ for stretching and
torsion. Since the vibration frequency is proportional to the
deformation energy, it makes sense that bending is preferred to
compression at small values of $k$ corresponding to long
wavelengths %
and large $H$ values, %
and vice versa for short wavelengths %
and small $H$ values.

As expanded upon further in the Appendix, we consider an elastic
tube of radius $R$ and height $H$ that could be either compressed
or bent by the displacement amplitude $A$. We will consider the
tube material to be described by the 2D elastic constant $c_{11}$
and the Poisson ratio $\alpha$. Then according to the equation
Eq.~(\ref{EqA4}) in the Appendix, we obtain for the total axial
compression energy
\begin{equation}
U_{c,tot} = %
\pi c_{11} \left(1-\alpha^2\right) R A^2 \frac{1}{H} \,. %
\label{eq27}
\end{equation}
Comparing this expression to Eq.~(\ref{eq24}), we can express
$c_{11}$ by
\begin{equation}
c_{11} = %
k \frac{H}{2{\pi}R(1-\alpha^2)} \,, %
\label{eq28}
\end{equation}
where $k$ is given by Eq.~(\ref{eq25}) and, for the sake of
simplicity, we use $R=R_o$.

According to Eq.~(\ref{eq1}),
assuming that load-induced changes of the stringer length $L$
can be neglected, %
any change in height $H$ would cause a reduction of the radius
$R$ and the circumference $2{\pi}R$.
We obtain
%
\begin{eqnarray}
\frac{{\delta}(2{\pi}R)}{2{\pi}R} &=& %
-\frac{H^2}{(4{\pi})^2 R^2} \frac{{\delta}H}{H} %
= - \frac{(H/R)^2}{16{\pi}^2} \frac{{\delta}H}{H}
\nonumber \\ %
&=&  - \alpha \frac{{\delta}H}{H} \,, %
\label{eq29}
\end{eqnarray}
%
thus defining the Poisson ratio
%
\begin{equation}
\alpha = \frac{(H/R)^2}{16{\pi}^2}\,. %
\label{eq30}
\end{equation}
%
According to Eq.~(%
\ref{EqA9}%
) of the Appendix, the total bending energy
is given by %
%
\begin{equation}
U_{b,tot} = %
4 {\pi}^5 c_{11} A^2 \left(\frac{R}{H}\right)^3
= 4 \pi^4 D_t \frac{A^2}{H^3} \,, %
\label{eq31}
\end{equation}
%
where $D_t={\pi}c_{11}R^3$ is the flexural rigidity of the tube.
The reduction in the height of the tube due to bending is given by
%
\begin{equation}
{\delta}H = \int_0^H  dx %
\left[ 1 + \left(\frac{du_z}{dx}\right)^2 \right]^{1/2} - H %
\approx \frac{2{\pi}^2A^2}{H} %
\label{eq32}
\end{equation}
%
to lowest order in $A$, and the work done by the external load is
thus %
%
\begin{equation}
F{\delta}H = \frac{2{\pi}^2FA^2}{H} \,.%
\label{eq33}
\end{equation}
%
The critical condition for bending to occur is that this work
should exceed the total bending energy, %
%
\begin{equation}
F{\delta}H > U_{b,tot} \,.%
\label{eq34}
\end{equation}
%
and translates to %
%
\begin{equation}
FH^2 > 2{\pi}^2D_t \,.%
\label{eq35}
\end{equation}
%
We can see from the parameters of the Loretto staircase that it is
very stable against buckling. From the above equations, we obtain
$2\pi^2D_t = 2\pi^3c_{11}R^3 = \pi^2R^2Hk/(1-{\alpha}^2)$, which
simplifies to $2\pi^2D_t = \pi^2R^2HEA/[H(1-{\alpha}^2)] =
\pi^2R^2EA/(1-{\alpha}^2)$. Since $R{\approx}R_o=1$~m,
${\alpha}=0.24$, and $EA=1.2{\times}10^8$~J/m, we get $2\pi^2D_t =
1.25{\times}10^9$~Jm. Assuming a compressive load $F=10^3$~N, this
quantity is vastly greater than $FH^2 = 10^3{\times}6.1^2 =
3.7{\times}10^4$~Jm. The load would have to be increased by more
than four orders of magnitude, or the height increased by more
than two orders of magnitude, to cause buckling. %

%
%


\section{Elastic Behavior of Similar Helical Structures in Nature}

Every helical structure, from the coiled telephone cord in
Fig.~\ref{fig1}(a) to the spiral staircase in Fig.~\ref{fig1}(b)
and to submicron-sized $\alpha$-helices found in proteins, can be
mapped topologically onto a helical coil. The helix we describe
here, which turns out very rigid, consists of two helical coils
with different radii, separated by a constant distance. This
particular design could clearly be utilized to form man-made
nanostructures that will be very rigid.

It is tempting to explore whether any existing structures in
Nature may look similarly and behave in a similar manner. Among
the biomolecules that immediately come to mind is the
double-stranded DNA that, coincidentally, is also left-handed.
DNA, however, does not fulfill the constant-step-width assumption,
since the bases from the two strands are non-covalently bound in
pairs, forming a ``breathing'' rather than a rigid unit. Another
system known for its toughness,
collagen~\cite{collagen-structure02}, has only some interstrand
covalent bonding, but not at every step. Moreover, its tripe-helix
structure differs from the model we discuss. After a long search,
we believe there are no real counterparts in Nature of the
structure we describe, at least not among biomolecules.


\section{Discussion}

Our main objective was to elucidate the origin of the previously
unexplained high rigidity of the unsupported spiral Loretto
staircase by developing a suitable formalism. Our numerical
results should be taken as rough estimates. We expect the local
axial deflections ${\delta}h$ of this staircase caused by load to
be significantly larger than the estimated values presented above.
The estimated value of the effective spring constant of the
staircase helix is likely to be reduced significantly by defects
and human-made joints in this all-wooden structure. Further
reduction would come from considering other deformation modes
including lateral compression or stretching of the wooden steps
and, to some degree, bending. Elastic response to shear stress in
the stringers should significantly contribute to the spring
constant especially in low-pitch spirals, with the coiled
telephone cord as an intuitive example. Even though the spiral
staircase of Fig.~\ref{fig1}(b) is a high-pitch spiral, allowing
for shear deformations in the stringers should further reduce its
effective force constant. Even if all these factors combined
should decrease the force constant by 1-2 orders of magnitude, we
may still expect a maximum local axial deflection ${\delta}h$ of
not more than $1-2$~cm in case that each of the 33 steps were
loaded by the weight of a person. As expanded above, since the
height is significantly larger than the radius, bending should not
play a significant role as a possible response to applied load. We
also note that at a later stage, the staircase had been augmented
by a railing, shown in Fig.~\ref{figA1}(a) in the Appendix. This
railing does not affect the elastic response of the staircase
under load.

As mentioned above, we have not found any asymmetric double-helix
structure in Nature that is rigid and does not stretch much.
Should such a structure exist, its stiffness should benefit from a
constant separation between the helical coils.


\section{Summary and Conclusions}

In summary, we have used continuum elasticity theory to describe
the elastic behavior of helical coils with an asymmetric
double-helix structure and have identified conditions, under which
they become very rigid. Theoretical insight gained for
macro-structures including a stretched telephone cord and an
unsupported helical staircase is universal and of interest for the
elastic behavior of helical structures on the micro- and nanometer
scale.


\section*{Acknowledgments}

A.G.E. acknowledges financial support by the South African
National Research Foundation Grant No.~80798. D.T. acknowledge
financial support by the NSF/AFOSR EFRI 2-DARE grant number
EFMA-1433459 and the hospitality of the University of the
Witwatersrand, South Africa, where this research was performed. We
thank Garrett B. King for his assistance with the literature
search for rigid helical structures in Nature and their schematic
graphical representation and thank Dan Liu for useful discussions.


\section*{Appendix}

\renewcommand\thesubsection{\Alph{subsection}}
\setcounter{subsection}{0} %
\renewcommand{\thefigure}{A\arabic{figure}}
\setcounter{figure}{0}
\renewcommand{\theequation}{A\arabic{equation}}
\setcounter{equation}{0} %

\subsection{Deformation Energy due to Compression and Bending}

As introduced in the main text, any helical structure may be
mapped onto an elastic tube of radius $R$ and height $H$ that
could be either compressed axially or bent. We will consider the
tube aligned along the $x$-direction and the tube material to be
described by the 2D elastic constant~\cite{DT260} $c_{11}$ and the
Poisson ratio $\alpha$. We will consider the local distortions to
be described by
\begin{equation}
u_x = (-A) \frac{x}{H} %
\label{EqA1}
\end{equation}
in the case of axial compression, and
\begin{equation}
u_z = %
A \left[ \sin{\left(2{\pi}\frac{x}{H}-\frac{\pi}{2}\right)} %
+ 1 \right] %
\label{EqA2}
\end{equation}
in the case of bending, where $A$ denotes the amplitude of the
distortion.

According to Eq.~(A2) of Reference [\onlinecite{DT260}], the
compression energy per length is given by
\begin{eqnarray}
U_c &=& \frac{1}{2} 2\pi R c_{11} \left(1-\alpha^2\right) %
    \left( \frac{du_x}{dx}\right)^2  \nonumber \\ %
    &=& \pi c_{11} \left(1-\alpha^2\right) R \left(\frac{A}{H}\right)^2 \,.
\label{EqA3}
\end{eqnarray}
%
\begin{figure}[b]
\includegraphics[width=1.0\columnwidth]{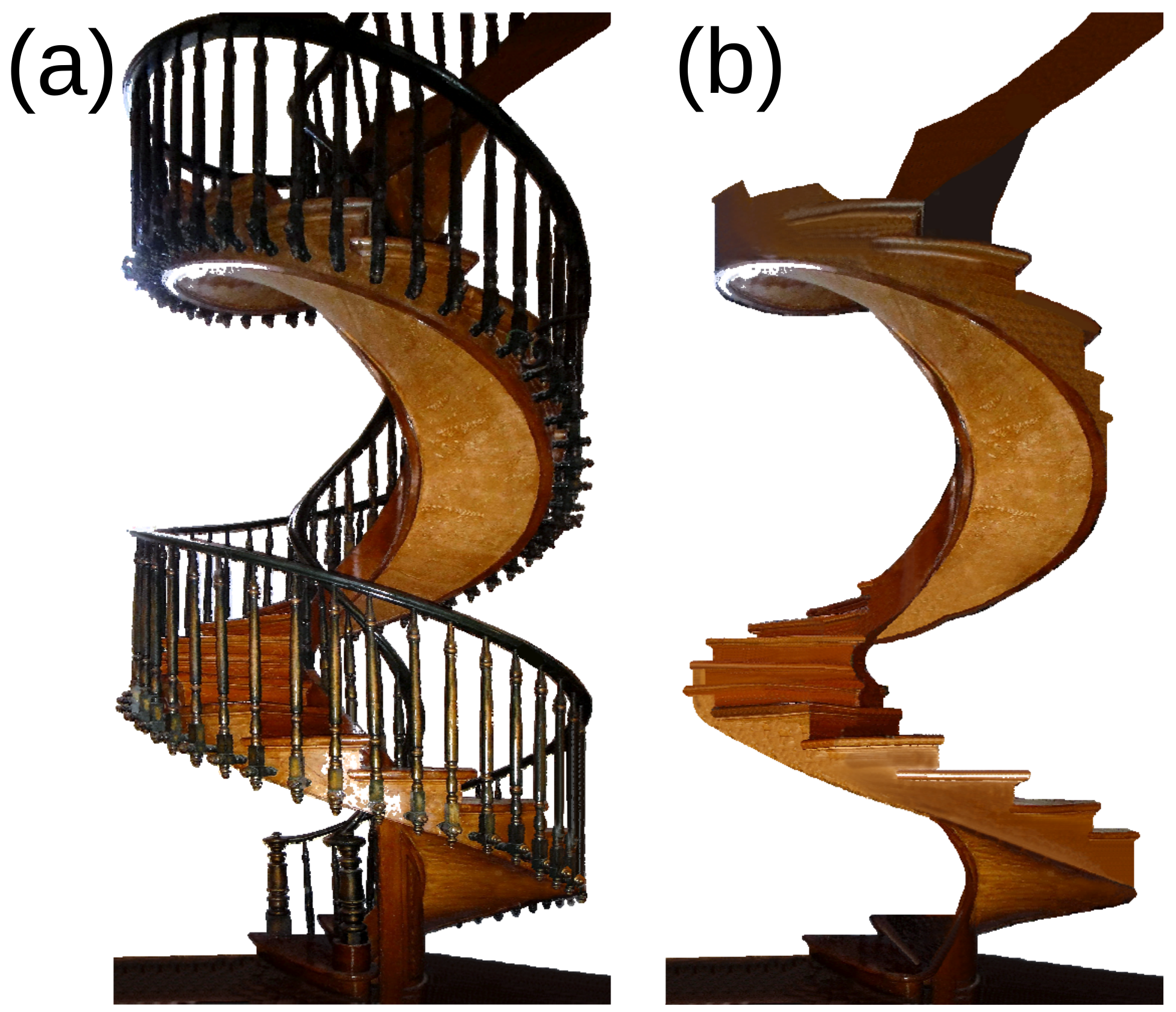}
\caption{Retouched photographs, with the background digitally
removed, of the spiral staircase in the Loretto Chapel, Santa Fe,
New Mexico. (a) Current view of the staircase including the
railing, which had been added long after construction. (b) Likely
view of the staircase as constructed, with no railing.
\label{figA1} }
\end{figure}

The total compression energy of the tube of height $H$ is then
\begin{equation}
U_{c,tot} = U_c H = %
\pi c_{11} \left(1-\alpha^2\right) R A^2 \frac{1}{H} \,. %
\label{EqA4}
\end{equation}
According to Eq.~(A15) of Reference [\onlinecite{DT260}], the
bending energy per length is given by
\begin{equation}
U_b =%
\frac{1}{2} ({\pi}c_{11}R^3 + {\pi}DR)%
\left(\!\frac{d^2{u_z}}{dx^2}\!\right)^2 \,.%
\label{EqA5}
\end{equation}
Since the flexural rigidity $D$ of the wall ``material'' vanishes
due to the separation between adjacent helix strands, we obtain
\begin{equation}
U_b = %
\frac{1}{2} {\pi} c_{11} R^3 \left(\frac{d^2{u_z}}{dx^2}\right)^2 \,.%
\label{EqA6}
\end{equation}
Using the expression in Eq.~(\ref{EqA2}) for the bending
deformation, we obtain
\begin{equation}
\left(\frac{d^2{u_z}}{dx^2}\right)^2 = %
A^2 \left(\frac{2\pi}{H}\right)^4 %
\sin^2{\left(2\pi\frac{x}{H}-\frac{\pi}{2}\right)} \,, %
\label{EqA7}
\end{equation}
which leads to
\begin{equation}
U_b = %
\frac{1}{2}{\pi}c_{11}R^3 A^2 %
\left(\frac{2\pi}{H}\right)^4 %
\sin^2{\left(2\pi\frac{x}{H}-\frac{\pi}{2}\right)} \,. %
\label{EqA8}
\end{equation}
The total bending energy is obtained by integrating $U_b$ in
Eq.~(\ref{EqA8}) along the entire height $H$ of the bent tube,
yielding
\begin{equation}
U_{b,tot} = %
4 {\pi}^5 c_{11} A^2 \left(\frac{R}{H}\right)^3 \,. %
\label{EqA9}
\end{equation}
Finally, assuming the same distortion amplitude $A$ for bending
and compression, we can determine the ratio of the compression and
the bending energy
\begin{equation}
\frac{U_{c,tot}}{U_{b,tot}} = %
\frac{1-\alpha^2}{4 {\pi}^4} \left(\frac{H}{R}\right)^2 %
\label{EqA10}
\end{equation}
that is independent of the amplitude $A$ and the elastic constant
$c_{11}$. We see that for $H>>R$, $U_{c,tot}>>U_{b,tot}$,
indicating that bending is energetically more affordable and thus
dominates. The opposite situation occurs for $H<<R$, when axial
compression dominates.


\subsection{Photographs of the Staircase}

Retouched photographs of the spiral staircase in the Loretto
Chapel, Santa Fe, New Mexico, are presented in Fig.~\ref{figA1}.



%

\end{document}